\documentstyle[epsf]{article}
\huge
\renewcommand{\theequation}{\arabic{section}.\arabic{equation}} 
\def\setzero{\setcounter{equation}{0}}

\newcounter{eqalph}
\def\bph{\setcounter{eqalph}{\value{equation}}
   \addtocounter{eqalph}{1}
   \setcounter{equation}{0}
   \renewcommand{\theequation}{\arabic{section}.\arabic{eqalph}\alph{equation}}}
\def\eph{\setcounter{equation}{\value{eqalph}}
   \renewcommand{\theequation}{\arabic{section}.\arabic{equation}}
\par\noindent}

\begin{document}

\baselineskip 18pt

\def \sech{{\rm sech}}
\def \tanh{{\rm tanh}}
\def \cn{{\rm cn}}
\def \sn{{\rm sn}}
\def\bm#1{\mbox{\boldmath $#1$}}
\newfont{\bg}{cmr10 scaled\magstep4}
\newcommand{\bzr}{\smash{\hbox{\bg 0}}}
\newcommand{\bzl}{%
   \smash{\lower1.7ex\hbox{\bg 0}}}
\title{Unitary Matrix Models  
     and  Phase Transition} 
\date{\today}
\author{ Masato {\sc Hisakado}  
\\
\bigskip
\\
{\small\it Department of Pure and Applied Sciences,}\\
{\small\it University of Tokyo,}\\
{\small\it 3-8-1 Komaba, Megro-ku, Tokyo, 113, Japan}}
\maketitle

\vspace{20 mm}

Abstract

We study  the   unitary matrix model with  a topological term.
We call the topological term the theta term.
In the symmetric model 
there is the  phase transition 
between the strong and weak coupling regime
at $\lambda_{c}=2$.
If the Wilson term is bigger  than 
the theta term,
 there is the strong-weak coupling  phase transition  
at the same $\lambda_{c}$.
On the other hand,
if the theta term is bigger   than the Wilson term,
there is only the strong coupling regime.
So the topological phase transition  disappears in this case.

\newpage

\section{Introduction}

Models  of the symmetric unitary matrix  are 
solved exactly in the double scaling limit,
 using orthogonal polynomials on  a circle.\cite{p}
The partition function is the form
$\int dU\exp\{-\frac{N}{\lambda}{\rm tr} (U+U^{\dag})\}$, 
where $U$ is an $N\times N$ unitary matrix.
We call the model symmetric model.\cite{b}

This unitary models has been studied in 
connection with the large-$N$ approximation 
to QCD in two dimensions.({\it one-plaquette model})\cite{gw}
Gross and Witten discovered  the third-order phase transition 
between the weak and strong coupling regime at $\lambda_{c}=2$.

We consider the model which has the 
 symmetric and 
anti-symmetric part.
The symmetric part is the usual Wilson action.
The anti-symmetric part becomes the topological term.\cite{kts}
We can see  the topological meaning 
of the theta term in the continuous limit.
 it gives rise to a  phase transition
 at $\theta=\pi$, 
if the Wilson term is bigger than the theta term.\cite{h3}
We call this  phase transition the topological phase transition and 
the model which has only the theta term 
the anti-symmetric model.

From the view point of the integrable system 
 this model can be embedded in the two-dimensional 
Toda hierarchy with the conjugate structure.\cite{m2},\cite{h2}
This Toda equation is split  into the modified Volterra (MV)  equation and the
discrete nonlinear Schr\"{o}dinger (DNLS) equation.
 MV and DNLS equations  correspond  to the 
symmetric and anti-symmetric model respectively. 
Coupling the Toda equation and the string equation
we can derive the third Painlev\'{e} (P III) equation.
We use P III to study the phase structure.

This letter is organized as follows.
In the section 2 we introduce the unitary matrix model with
 a topological term.
We consider the two cases:

(i)  the Wilson  term is bigger   than the theta  term,

(ii) the theta term is bigger  than the theta term.

In the section 3 we consider the phase transition in the case (i)
using the P III.
In the section 4 we study the phase structure in the case (ii).
The last section is devoted to the concluding remarks.

\setzero

\section {Wilson term and  theta term}
We consider  the partition function 
of the  unitary matrix model with  a topological term.
We consider the unitary matrix model 
\begin{equation}
Z_{N}=\int dU\exp (-\frac{N}{\lambda}V(U)),
\end{equation}
where $V(U)$ is a potential
\begin{equation}
V(U)=t_{1}U+t_{-1}U^{\dag}.
\label{pot}
\end{equation}
$U$ is the gauge group $U(N)$.

We  divide   the potential  into the symmetric and the anti-symmetric
 part, 
\begin{equation}
V(U)=t_{+}s_{w}+t_{-}s_{\theta},
\end{equation}
where
\begin{equation}
t_{+}
=
\frac{t_{1}+t_{-1}}{2},\;\;\;
t_{-}
=
\frac{t_{1}-t_{-1}}{2}.
\end{equation}
$s_{w}$ is the symmetric part, the usual Wilson action
\bph
\begin{equation}
s_{w}=\frac{1}{2}({\rm tr}U+{\rm tr}U^{\dag}).
\end{equation}
Here we choose 
\begin{equation}
s_{\theta}=\frac{1}{2}({\rm tr}U-{\rm tr}U^{\dag}),
\end{equation}
\eph
for the theta term, the anti-symmetric part.

Hereafter we call next reduced models  
the symmetric model and the anti-symmetric model:
\begin{eqnarray*}
t_{1}&=&t_{-1}=t_{+}, \;\;\;t_{-}=0,\;\;\;\;\;({\rm symmetric\;\;model})
\\
t_{1}&=&-t_{-1}=t_{-}, \;\;\;t_{+}=0.\;\;\;\;\;({\rm anti-symmetric \;\;model}) 
\end{eqnarray*}
\eph

Here we consider the two cases 
\begin{equation}
{\rm (i) }|t_{+}| > |t_{-}|,\;\;\;\;\;\;\;\; 
{\rm (ii)} |t_{-}| >|t_{+}|.
\nonumber 
\end{equation}
We parameterize $t_{1}$ and $t_{-1}$ by  $\epsilon$:
\begin{eqnarray}
&{\rm (i)}&  t_{1}=-e^{\epsilon},\;\;\;t_{-1}=-e^{-\epsilon},
\nonumber \\
&{\rm (ii)}&
 t_{1}=-e^{\epsilon},\;\;\;t_{-1}=e^{-\epsilon}.
\end{eqnarray}

The measure $dU$ may be written as 
\begin{equation}
dU=\prod_{m}^{N}\frac{d\alpha_{m}}{2\pi}
\Delta(\alpha)\bar{\Delta}(\alpha).
\end{equation}

 Here the eigenvalues of $U$ 
 are $\{\exp(i\alpha_{1}),\exp(i\alpha_{2}),\cdots,\exp(i\alpha_{N})\}$
and 
$\Delta\bar{\Delta}$
 is the Jacobian for the change of variables,
\begin{eqnarray}
\Delta(\alpha)&=&{\rm det}_{jk}e^{i\alpha_{j}(N-k)},
\nonumber \\ 
\bar{\Delta}(\alpha)&=&{\rm det}_{jk}e^{-i\alpha_{j}(N-k)}.
\end{eqnarray}

Then 
we obtain the partition function in  the case (i):
\begin{equation}
Z_{N}={\rm const.}
{\rm det}_{jk}
e^{\epsilon(-j+k)}I_{-j+k}(N/\lambda)={\rm const.}{\rm det}_{jk}
I_{-j+k}(N/\lambda).
\label{pf21}
\end{equation}
Here $I_{-j+k}$ is the modified Bessel function of order $-j+k$.
In the same way 
we can calculate the partition function in  the case (ii):
\begin{equation}
Z_{N}={\rm const.}
{\rm det}_{jk}
e^{\epsilon(-j+k)}J_{-j+k}(N/\lambda)
={\rm const.}
{\rm det}_{jk}
J_{-j+k}(N/\lambda).
\label{pf22}
\end{equation}
Here $J_{-j+k}$ is the Bessel function of order $-j+k$.
Notice that (\ref{pf21}) and (\ref{pf22}) 
 do  not depend on $\epsilon$.

It is well known that  the partition function $Z_{N}$ 
of the unitary matrix model  can be presented  as 
a product  of norms  of the biorthogonal polynomial system.
Namely, let us introduce  a scalar product of the form 
\begin{equation}
<A,B>=\oint\frac{d\mu(z)}{2\pi i z}
\exp\{-V(z)\}
A(z)B(z^{-1}),
\end{equation}
where 
\begin{equation}
V(z)=t_{1}z+t_{-1}z^{-1}.
\end{equation}
Let us define  the system of the polynomials  biorthogaonal 
with respect to this scalar product 
\begin{equation}
<\Phi_{n},\Phi_{k}^{*}>=h_{n}\delta_{nk}.
\label{or}
\end{equation}
Then, the partition function  $Z_{N}$ 
is equal to the product of $h_{n}$'s:
\begin{equation} 
Z_{N}=\prod_{k=0}^{N-1}h_{k},\;\;\;Z_{0}=1.
\end{equation}
The polynomials are normalized as follows
(Note that the asterisk  `*' does not mean the 
complex conjugation):
\begin{equation} 
\Phi_{n}=z^{n}+\cdots+S_{n-1},\;\;\Phi_{n}^{*}
=z^{n}+\cdots+S_{n-1}^{*},\;\;
S_{-1}=S_{-1}^{*}\equiv 1.
\label{2.4}
\end{equation}
Now it is easy to show that these polynomials satisfy the following
recurrent relations, 
\begin{eqnarray}
\Phi_{n+1}(z)&=&z\Phi_{n}(z)+S_{n}z^{n}\Phi_{n}^{*}(z^{-1}),
\nonumber \\
\Phi_{n+1}^{*}(z^{-1})&=&z^{-1}\Phi_{n}^{*}(z^{-1})
+S_{n}^{*}z_{-n}\Phi_{n}(z),
\end{eqnarray}
and
\begin{equation}
\frac{h_{n+1}}{h_{n}}=1-S_{n}S_{n}^{*}.
\end{equation}

From (\ref{or}) we can obtain the 
string equations:
\bph
\begin{equation}
 (n+1)S_{n}=(t_{-1}S_{n+1}+t_{1}S_{n-1})
(1-S_{n}S_{n}^{*}),
\label{edp3}
\end{equation}
\begin{equation}
 (n+1)S_{n}^{*}=(t_{1}S_{n+1}^{*}+t_{-1}S_{n-1}^{*})
(1-S_{n}S_{n}^{*}).
\label{edp4}
\end{equation}
\eph

In the unitary matrix model
there is a  conjugate relation:
\begin{equation}
t_{1}S_{n}S_{n-1}^{*}=t_{-1}S_{n}^{*}S_{n-1}.
\label{2}
\end{equation}
Here we define  $a_{n}$:
\begin{equation}
a_{n}\equiv 1--S_{n}S_{n}^{*}=\frac{h_{n+1}}{h_{n}}.
\end{equation}
{}From (\ref{2})  $a_{n}$ are  functions of the radial coordinate
\begin{equation}
x=t_{1}t_{-1},
\label{rc}
\end{equation}
only.

$a_{n}$ satisfies the next Painlev\'{e} V with $\delta_{V}=0$:\cite{h2}
\begin{eqnarray}
\frac{\partial ^{2}a_{n}}{\partial x^{2}}
&=&
\frac{1}{2}(\frac{1}{a_{n}-1}+\frac{1}{a_{n}})
(\frac{\partial a_{n}}{\partial x})^{2}
-\frac{1}{x}\frac{\partial a_{n}}{\partial x}
\nonumber \\
& &
-\frac{2}{x}a_{n}(a_{n}-1)
+\frac{(n+1)^{2}}{2x^{2}}
\frac{a_{n}-1}{a_{n}}.
\label{p3}
\end{eqnarray}

\setzero
\section{Phase structure in  the case (i)}

The partition function 
dose not depend on $\epsilon$ from (\ref{pf21}).
This  can be seen from the radial coordinate (\ref{rc}).
From these results in the large-$N$ limit the phase structure 
in  the case (i) is the same as the symmetric model.\cite{gw}
To study the strong-weak coupling phase transition 
we use (\ref{p3}) in $x\rightarrow \infty$.
We rewrite  (\ref{p3}) 
a second order ODE  for $S_{n}$
\begin{equation}
\frac{\partial^{2} S_{n}}{\partial t_{+}^{2}}
=
-\frac{S_{n}}{1-S_{n}^{2}}
(\frac{\partial S_{n}}{\partial t_{+}})^{2}
-\frac{1}{t_{+}}\frac{\partial S_{n}}{\partial t_{+}}
+\frac{(n+1)^{2}}{t_{+}^{2}}
\frac{S_{n}}{1-S_{n}^{2}}
-4S_{n}(1-S_{n}^{2}),
\label{piii}
\end{equation}
where $t_{+}=N/\lambda$.
This equation can be obtained directly 
from coupling 
the string equation and the modified Volterra equation.\cite{h2}

In particular when we consider the strong coupling regime
  we only need  to
solve
\begin{equation}
\frac{\partial S_{n}}{\partial t_{+}^{2}}
+\frac{1}{t_{+}}\frac{\partial S_{n}}{\partial t_{+}}
-[\frac{(n+1)^{2}}{t_{+}^{2}}-4]S_{n}=0.
\end{equation}
This is the Bessel equation.
Then setting $n=N$, we can obtain in the large-$N$ limit
\begin{equation}
S_{N}=J_{N}(2t^{+})+O(\frac{1}{\lambda^{(3N+2)}})
\stackrel{N\rightarrow\infty}{\longrightarrow}
J_{N}(2N/\lambda),
\end{equation}
where 
$J_{N}$ is the standard Bessel function.
As a consequence, we find
\begin{equation}
S_{N}\sim
\exp N[\sqrt{1-\frac{4}{\lambda^{2}}}
-\log \frac{\lambda(1+\sqrt{1-4/\lambda^{2}}}{2})].
\end{equation}

(\ref{piii}) is especially appropriate 
for a discussion  of the weak coupling regime.
One may consider an $1/N$ expansion for $S_{N}$:\cite{gl}
\begin{equation}
S_{N}\rightarrow
\sqrt{1-\frac{\lambda}{2}}
-\frac{1}{N^{2}}\frac{\lambda^{3}}{128} 
(1-\frac{\lambda}{2})^{5/2}+O(\frac{1}{N^{4}}).
\end{equation}
We notice the typical phase transition
behavior  of $S_{N}$.
The critical point $\lambda_{c}=2$
 is independent of 
the parameter $\epsilon$.

\setzero

\section{Phase structure in  the case (ii)}

In the large $N$-limit to study the phase structure
we use the string equations (\ref{edp3}) and 
(\ref{edp4}).
Setting $n=N$, there is a critical point 
at $\lambda_{c}=2\sinh \epsilon$ 
for the roots of the differential 
equation are degenerate.
It seems that 
in the limit $\epsilon\rightarrow 0$
 (the Wilson term vanishes)
the weak coupling regime disappears. 
But as seeing in the previous section this is not correct.
From the radial coordinate  or (\ref{pf22})
the critical point
does not depend on $\epsilon$.
The phase structure in  the case (ii) is the same as 
the anti-symmetric model.
To see the phase structure we study (\ref{p3}) 
in $x\rightarrow -\infty$.
Then 
the difference betweeen  the case (i) and (ii) is 
 the sign of the third term of RHS of (\ref{p3}).
We rewrite  (\ref{p3}) 
a second order ODE  for $S_{n}$:
\bph
\begin{eqnarray}
\frac{\partial^{2} S_{n}}{\partial t_{-}^{2}}
=
-\frac{S_{n}}{1-S_{n}^{2}}
(\frac{\partial S_{n}}{\partial t_{-}})^{2}
-\frac{1}{t_{-}}\frac{\partial S_{n}}{\partial t_{-}}
+\frac{(n+1)^{2}}{t_{-}^{2}}
\frac{S_{n}}{1-S_{n}^{2}}
&+&4S_{n}(1-S_{n}^{2}),
\nonumber \\
& &
(n={\rm odd})
\label{piii1}
\end{eqnarray}
and 
\begin{eqnarray}
\frac{\partial^{2} S_{n}}{\partial t_{-}^{2}}
=
-\frac{S_{n}}{1+S_{n}^{2}}
(\frac{\partial S_{n}}{\partial t_{-}})^{2}
-\frac{1}{t_{-}}\frac{\partial S_{n}}{\partial t_{-}}
+\frac{(n+1)^{2}}{t_{-}^{2}}
\frac{S_{n}}{1+S_{n}^{2}}
&+&4S_{n}(1+S_{n}^{2}),
\nonumber \\
& &
(n={\rm even})
\label{piii2}
\end{eqnarray}
\eph
using
\begin{eqnarray}
S_{n}&=&S_{n}^{*},\;\;\;\;a_{n}=1-S_{n}^{2},\;\;\;\;\;\;\;(n={\rm odd}),
\nonumber \\
S_{n}&=&-S_{n}^{*},\;\;\;\;a_{n}=1+S_{n}^{2},\;\;\;\;\;\;\;(n={\rm even}).
\label{sid}
\end{eqnarray}
Notice $t_{-}=2N/\lambda$.
This equation can be obtained directly from the coupling 
the string equations and the 
 discrete nonlinear Schr\"{o}dinger equation.\cite{h2}
When we consider the strong coupling regime we only need to solve
\begin{equation}
\frac{\partial S_{n}}{\partial t_{-}^{2}}
+\frac{1}{t_{-}}\frac{\partial S_{n}}{\partial t_{-}}
-[\frac{(n+1)^{2}}{t_{-}^{2}}+4]S_{n}=0.
\end{equation}
This is the modified Bessel equation.
Then setting  $n=N$, we can obtain  in the large-$N$ limit 
\begin{equation}
S_{N}=K_{N}(2t_{-})+O(\frac{1}{\lambda^{(3N+2)}})
\stackrel{N\rightarrow\infty}{\longrightarrow}
K_{N}(2N/\lambda),	
\end{equation}
where 
$K_{N}$ is the second kind modified  Bessel function.
As a consequence,  we obtain 
\begin{equation}
S_{N}\sim \exp N\sqrt{1+\frac{4}{\lambda^{2}}}.
\end{equation}
On the other hand we can not do an $1/N$ expansion  in (\ref{piii1}) and 
(\ref{piii2}), since there is not the weak  coupling regime.
Then there is not the strong-weak  coupling phase transition in the case (ii).

In  the previous letter, \cite{h3}
adding  the term $l\log U$
 we have shown that 
there is the phase transition 
at $\theta=\pi$ in the case (i).
We call the phase transition the topological phase transition.
This phase transition can be seen in the weak  coupling regime.\cite{kts}
So in the case (ii) 
there is not
the topological phase transition, too.

\setzero
\section{Concluding remarks}
We study the unitary matrix model with a topological term.
This model contains the Wilson term and the theta term.
The Wilson term is the  symmetric  and the theta  is the 
anti-symmetric part.
We call the model which has only the Wilson term 
the symmetric model and the model
which has only the theta term the anti-symmetric model.
It is well known that in the symmetric model 
there is the  strong-weak  coupling  phase transition  at $\lambda_{c}=2$.
If the Wilson term is bigger  than the theta term, 
there is the strong-weak coupling phase transition  at the same $\lambda_{c}$.
Adding  the term $l\log U$,
 there is the topological phase transition at $\theta=\pi$.
On the other hand if the theta term is bigger  than the Wilson term, 
there is no   strong-weak  coupling phase transition.
 Since there is only the strong coupling  regime,
 the topological phase transition  also disappears.

\end{document}